\documentclass[reprint,aps,twocolumn,superscriptaddress,floatfix,preprintnumbers,amsmath,amssymb]{revtex4-1}

\usepackage{graphics,amssymb,amsmath,epsfig,color}
\usepackage{graphicx}   
\usepackage{subfigure}
\usepackage{float}
\usepackage{dcolumn}   
\usepackage{bm}        
\usepackage{amssymb}   
\usepackage{epstopdf}
\usepackage{blindtext}
\usepackage{filecontents}

\let\oldcite=\cite 
\renewcommand{\cite}[1]{\textcolor[rgb]{0,0,1}{\oldcite{#1}}}

\let\oldref=\ref 
\renewcommand{\ref}[1]{\textcolor[rgb]{0,0,1}{\oldref{#1}}}

\begin{document}

\title{ Defective annular semiconductor-superconductor photonic crystal  }

\author{Alireza Aghajamali}
\email{alireza.aghajamali@curtin.edu.au }
\affiliation{Department of Physics and Astronomy, Curtin University, Perth, Western Australia 6102, Australia}
\author{Tannaz Alamfard}
\affiliation{Department of Optics and Laser Engineering, Marvdasht Branch, Islamic Azad University, Marvdasht, Iran}
\affiliation{Young Researchers Club, Marvdasht Branch, Islamic Azad University, Marvdasht, Iran}


\begin{abstract}
In this paper, the transfer matrix method is used to investigate the reflectance properties in a symmetric defective annular photonic crystal (APC) containing semiconductor, high-$T_c$ superconductor and a radial defect layer. Numerical results offer many noteworthy features that can be very useful in designing optical devices. In this regard, the geometric effects of alternate layers on the optical reflectance of our defective APC structure by changing the thickness of different layers would be investigated. Then, the possibility of controlling the spectral position of existed photonic band gaps (PBGs) and defect modes would be discussed by our analysis. In addition, it is found that the characteristics of the reflectance spectrum and subsequently the spectral position of PBG and defect mode are entirely independent on changes in the starting radius of the hollow core. This characteristic is very attractive to manufacturers, since increasing the size of the core radius would offer flexibility and simplicity of fabrication in producing optical devices. This study reveals that the optical reflectance of defective APCs is strongly dependent on azimuthal mode number. Therefore, our designed defective APC structure works as a perfect reflector for the considerable range of wavelength for higher azimuthal mode numbers. Lastly, our results show that by increase the temperature of superconductor layer both of the existed band gaps and defect mode show a red-shift trend by moving towards higher wavelengths. The proposed structure and related results can lead to gain valuable information for designing and fabrication of new types of annular Bragg resonators surrounding a radial defect and integrated visible waveguide devices like optical switches and filters. 

\end{abstract}


\date{\today}

\maketitle

\section{Introduction}
\label{Intro}

The study of electromagnetic waves propagation in dielectric stratified construction has been conducted for a long time \cite{yeh1977electromagnetic, *yeh1988optical}. Since the concept of the photonic crystals (PCs), a multilayer optical structure consisting of two or more periodic arrangement of materials with different dielectric constants, was presented by two novel works of Yablonovitch and John, this subject has been attracted a lot of attention by the electromagnetic and optical communities again in 1987 \cite{yablonovitch1987inhibited, *PhysRevLett.58.2486}. As the electromagnetic wave propagates in PCs, it is prohibited in a specified span of frequencies (or wavelengths) named photonic band gaps (PBGs). PCs have acquired many practical applications in modern photonic engineering, including optoelectronics and optical communications because of their capability to direct and regulate the dispersion of light \cite{yablonovitch1987inhibited,*PhysRevLett.58.2486, weiss2005tunable, srivastava2009broadband, aghajamali2014effects,*aghajamali2012effects, xu2005design, *ren2006photonic, *mekis1996high, nayak2017tunable, *nayak2020dodecanacci-arxiv, *nayak2020photonic, zimmermann2004photonic, srivastava2014investigation, biswal2020photonic, solaimani2020optical,*nayak2020dodecanacci,*solaimani2020band, *nayak2017effect,*nayak2019robust}.

The initial PBG structures were significantly created by placing common dielectrics, metals, and semiconductors in a periodic multilayer structure. However, some investigations into the PBGs in a PCs composed of superconducting and dielectric materials have been conducted \cite{takeda2004properties, ooi1999polariton, *ooi2000photonic, *wu2005photonic, berman2006superconducting}. There are some significant reported differences between this kind of superconducting planar Bragg reflector (SPBR) and an all-dielectric plane Bragg reflector. For instance, incorporating superconducting materials and combined effects of alternation would contribute to a low-frequency PBG \cite{ooi1999polariton, *ooi2000photonic, *wu2005photonic}.

When the periodicity of a simple PC is broken, such as altering the thickness of a layer \cite{boedecker2003all}, or inserting another medium into the structure \cite{akahane2003high}, some defective modes will be produced within the PBGs. In a conventional PC of $(AB)_{n}$, by adding a defected layer known as $D$ a defective structure like  $(AB)_{n}D(AB)_{n}$ will be produced, which in turn results in generating defect mode in the PBGs. The defect mode which is also called resonant transmission peaks would be observed in the transmission spectrum with a pointed narrow resonant peak owing to the change of the interference behavior of the light. The structure of the defect modes in transmittance is mostly utilized to design a PC-based narrowband transmission filter \cite{Orfanidis}, and such kinds of practical application of defective PCs has made this structure appealing to researchers for conducting further investigations \cite{aghajamali2014defect, *aghajamali2015analysis, *barati2016near, *aghajamali2016near, *nayak2017double}.

The geometrical structures of the photonic crystals are categorized into one-dimensional (1D), two-dimensional (2D), and three-dimensional (3D), based on the direction of modulation in refractive index. The fabrication of the 1D PC is considered easier than 2D and 3D photonic crystals due to the possibility of more production at any wavelength. Moreover, 1D PCs can be applied to survey many essential optical features, such as the characteristics of omnidirectional mirror and the existence of PBGs \cite{fink1998dielectric, winn1998omnidirectional}. In 1D PCs the wave propagation features would be analyzed based on the usual transfer matrix method (TMM) in Cartesian coordinates engendered by Abeles theory \cite{yeh1977electromagnetic, yeh1988optical, born1999fraunhofer}.

As well as the typical planar 1D PC, wave propagation in a cylindrical multilayer structure has also attracted much attention recently by scientific communities \cite{jebali2007analytical, ochoa2000diffraction, urzhumov2010transformation, liang2011scaling, toda1990single, fallahi1992electrically, green2004vertically, scheuer2003two, scheuer2004low}. A cylindrical Bragg reflector or cylindrical photonic crystal (CPC) is a kind of annular photonic crystal (APC) with an alternate cylindrical multilayer structure. The annular periodical structure has recently utilized in different approaches in science such as conformal mapping \cite{heiblum1975analysis}, coupled mode theory \cite{erdogan1992circularly, scheuer2003coupled}, transfer matrix method \cite{ping1994transmission}, and finite-difference and time domain approach.

Like in a planar PC, another version of TMM, which was recently developed by Kaliteevski et al.~\cite{kaliteevski1999bragg}, would be applied to analyze the electromagnetic wave propagation characteristics in a CPC accurately. As a matter of fact, this well-developed TMM in cylindrical coordinates is a similar version of Abeles theory of TMM in Cartesian coordinates. Cylindrical TMM provides a context for exploring the reflection features for the CPC and consequently performing a comparison with planar 1D PC \cite{chang2011investigation}. Besides, cylindrical TMM prepares the ground for the investigation of photonic band structures in superconducting and metallic CPCs \cite{chen2012narrowband, *chen2009optical}. By advancing modern manufacturing technology, annular Bragg lasers (or resonators) consist of ABRs surrounding a radial defect have been developed \cite{green2004vertically, scheuer2003two, scheuer2004low, scheuer2003annular}. 

Such types of resonators have a specific characteristic of vertical emission, which creates them an atmosphere to be used in optical communication and optoelectronics \cite{scheuer2004annular}. In addition, they have more precise measurement ability in the biological and chemical functions \cite{scheuer2003annular} than common ones, as well as practical use in optical communication system. Similar to the planar PCs that can composed on different materials such as metamaterial, semiconductor metamaterials, magnetized cold plasma and semiconductor \cite{aghajamali2016transmittance, srivastava2017analysis, aly2016analysis, nayak2019near}, various type of CPCs can generate with composition of different materials \cite{el2017optical, el2017photonic, roshan2017optical, abadla2020thermo, chen2009wave, srivastava2016study, srivastava2016investigation}. Chen et al. \cite{chen2009optical}  have carried out an investigation into the optical features of a superconducting annular bilayer photonic crystal in their previous research. 

In this work, we consider the superconducting media as one of the layers due to having some essential properties such as lower dispersion, lower dielectric losses, and wider bandwidths in comparison with usual metals. Multilayer structure containing superconductors and dielectric medium have also received considerable attention recently \cite{ooi2010echo, aly2009thz, srivastava2014study}. Transfer matrix method for the cylindrical waves, which was developed by the Kaliteevski et al. \cite{kaliteevski1999bragg}, will be applied to acquire the reflectance or transmittance spectrum of the APC structure. We will investigate the effects of azimuthal mode number (\textit{m}), starting radius ($\rho_{0}$), temperature, and thickness of the defect layer on the defect modes of a defective annular photonic crystal by studying the reflectance spectra. It is found that by increasing the related parameters, the number and wavelength of the defect modes would change.

\section{Theoretical Modeling and Formulation}

\begin{figure}[b]
\centering\includegraphics[width=\linewidth]{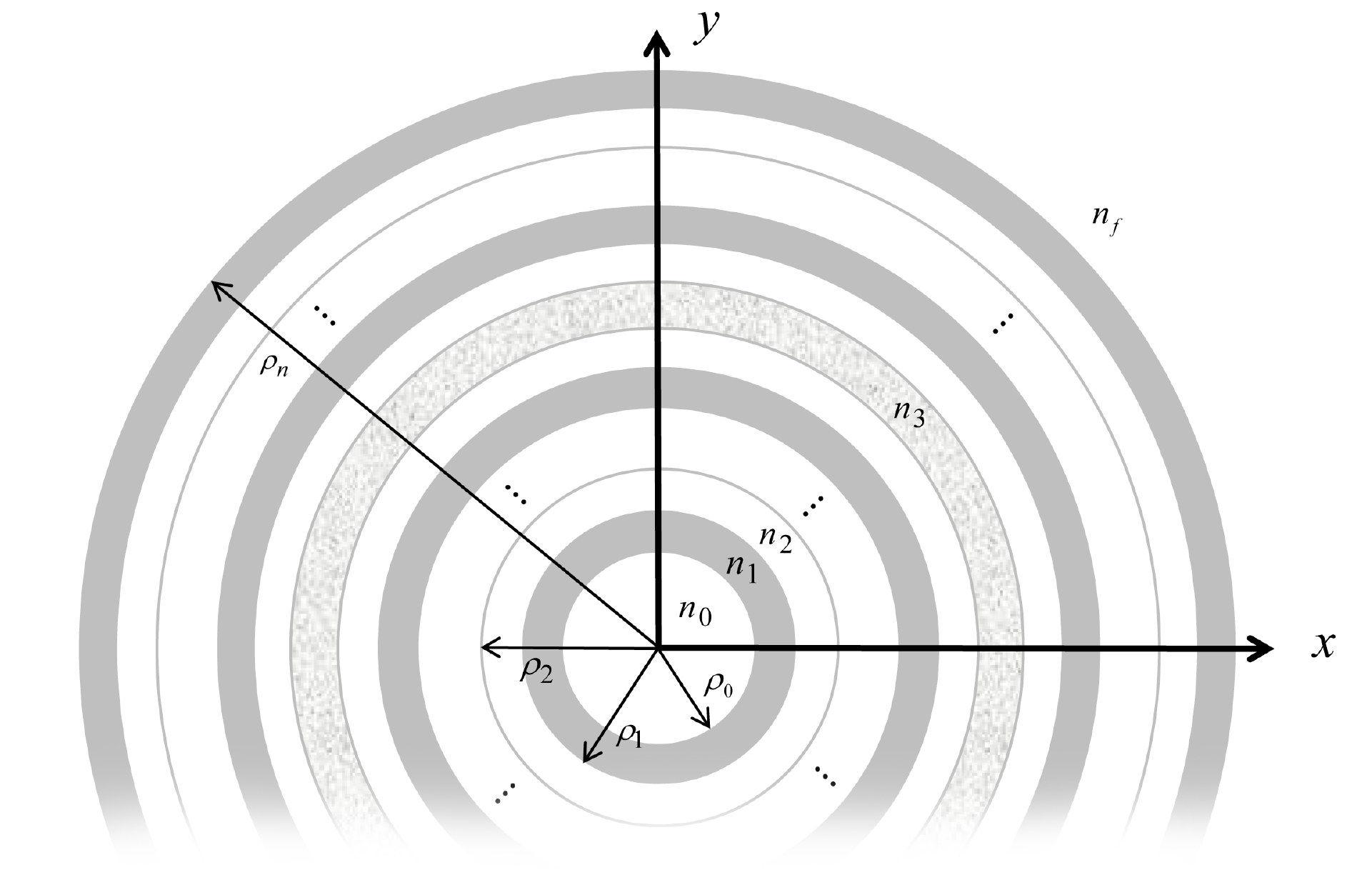}
\caption{ The cross sectional view of a hollow core defective APC immersed in free space i.e. $n_{0}=n_{f}=1$, where periodic layers including semiconductor and superconductor are taken to be with indices $n_1$ and $n_2$, and radii $\rho_{1}$ and $\rho_{2}$, respectively. A symmetrical radial defect layer with refractive index $n_3$ is located at the middle of the periodic layers of the cylinder. }
\label{fig1}
\end{figure}

A defective annular photonic crystal is modeled as a periodic multilayered structure immersed in free space, where the structure of periodic layers are interrupted with a defect layer at the middle of the cylinder, as illustrated in Figure~\ref{fig1}. The cylindrical structure is composed of an innermost section, where the refractive index and starting radius of the starting medium are $n_{0}$ and $\rho_{0}$, respectively. The periodic layers are considered as semiconductor (silicon monoxide, SiO) and high-temperature superconductor Bi$_{2}$Sr$_{2}$CaCu$_{3}$O$_{8}$~(BSCCO) with refractive indices $n_{1}$ and $n_{2}$, radii $\rho_{1}$ and $\rho_{2}$, and thickness $d_{A}$ and $d_{B}$, respectively. We assume that the middle section of the cylinder is occupied symmetrically by a radial defect layer of (Indium nitride, InN) with refractive index $n_{3}$ and thickness $d_{C}$. The outermost section (last layered medium) of the cylindrical structure has refractive index $n_{f}$ and radius $\rho_{f}$. Moreover, the total number of lattice period of the structure is denoted by \textit{N}. It is supposed that the cylindrical wave would diverge from the axis of symmetry $\rho=0$ and then interfere normally with the first junction at $\rho=\rho_{0}$. The transfer matrix method is applied to analyze the reflectance at the first circular boundary $\rho=\rho_{0}$ in the cylindrical Bragg wave \cite{kaliteevski1999bragg, chen2009optical}.

Supposing that the time dependent part of all the electromagnetic fields is $\mathrm{exp}(j \omega t)$, the source-free two curl Maxwell's equations are written as
\[
\nabla \times E=-j \omega \mu H
\]
\[
\nabla \times H=j \omega \varepsilon E
\]

There are two possible modes (or polarizations) including TE and TM modes for the cylindrical Bragg wave. $E_Z$, $H_{\phi}$, and $H_{\rho}$ are the non-zero fields, which satisfy the following three equations in each layer for TE mode:

\begin{equation}
\frac{1}{\rho} \frac{\partial E_{Z}}{\partial \phi}=-j \omega \mu H_{\rho} 
\label{equ1}
\end{equation}
\begin{equation}
\frac{\partial E_{Z}}{\partial \rho}=j \omega \mu H_{\phi}
\label{equ2}
\end{equation}
\begin{equation}
\frac{\partial\left(\rho H_{\phi}\right)}{\partial \rho}-\frac{\partial H_{\rho}}{\partial \phi}=j \omega \varepsilon \rho E_{Z}
\label{equ3}
\end{equation}

The governing equation for $E_Z$ can be derived by making use of above equations and is given by

\begin{equation}\small
\rho \frac{\partial}{\partial \rho}\left(\rho \frac{\partial E_{Z}}{\partial \rho}\right)-\rho^{2} \frac{1}{\mu} \frac{\partial \mu}{\partial \rho} \frac{\partial E_{Z}}{\partial \rho}+\frac{\partial}{\partial \phi}\left(\frac{\partial E_{Z}}{\partial \phi}\right)+\omega^{2} \mu \varepsilon \rho^{2} E_{Z}=0
\label{equ4}
\end{equation}

The solution for $E_Z$ would be achieved by applying the method of separation of variables and is stated as

\begin{equation}\small
E_{Z}(\rho, \phi)=V(\rho) \varphi(\phi)=\left[A J_{m}(k \rho)+B Y_{m}(k \rho)\right] \exp (j m \phi)
\label{equ5}
\end{equation}
where $J_{m}$, $Y_{m}$ and \textit{m} are respectively a Bessel function, Numann function and the azimuthal number. \textit{A} and \textit{B} are considered as constant values, and $k=\omega \sqrt{\mu \varepsilon}$ is the wave number of the related material. The azimuthal part of the magnetic field $H_{\varphi}$ by making use of Equation~\ref{equ2} is given by

\begin{equation}\small
H_{\phi}(\rho, \varphi)=U(\rho) \phi(\varphi)=-j p\left[J_{m}^{\prime}(k \rho)+B Y_{m}^{\prime}(k \rho)\right] \exp (j m \varphi)
\label{equ6}
\end{equation}

In the above mentioned-equation $p=\sqrt{\mu / \varepsilon}$ is the intrinsic admittance of the medium. A single layer matrix relation corresponding to the electric and magnetic fields at its two interfaces can be obtained according to the equations (5) and (6). The related matrix for the first layer $M_1$, with refractive index $n_1$ and interfaces at $\rho=\rho_{0}$ and $\rho=\rho_{1}$ would be expressed as \cite{kaliteevski1999bragg}

\begin{equation}
\left( \begin{array}{l}{V\left(\rho_{1}\right)} \\ {U\left(\rho_{1}\right)}\end{array}\right)=M_{1} \left[ \begin{array}{c}{V\left(\rho_{0}\right)} \\ {U\left(\rho_{0}\right)}\end{array}\right]
\label{equ7}
\end{equation}

where the single-layer matrix  is given as,
\[
M_{1}=\left[ \begin{array}{ll}{m_{11}} & {m_{12}} \\ {m_{21}} & {m_{22}}\end{array}\right]
\]

The elements of matrix $M_1$ are considered as,
\[\small
m_{11}=\frac{\pi}{2} k_{1} \rho_{0}\left[Y_{m}^{\prime}\left(k_{1} \rho_{0}\right) J_{m}\left(k_{1} \rho_{1}\right)-J_{m}^{\prime}\left(k_{1} \rho_{0}\right) Y_{m}\left(k_{1} \rho_{1}\right)\right]
\]
\[\small
m_{12}=j \frac{\pi}{2} \frac{k_{1}}{p_{1}} \rho_{0}\left[J_{m}\left(k_{1} \rho_{0}\right) Y_{m}\left(k_{1} \rho_{1}\right)-Y_{m}\left(k_{1} \rho_{0}\right) J_{m}\left(k_{1} \rho_{1}\right)\right]
\]
\[\small
m_{21}=-j \frac{\pi}{2} k_{1} \rho_{0} p_{1}\left[Y_{m}^{\prime}\left(k_{1} \rho_{0}\right) J_{m}^{\prime}\left(k_{1} \rho_{1}\right)-J_{m}^{\prime}\left(k_{1} \rho_{0}\right) Y_{m}^{\prime}\left(k_{1} \rho_{1}\right)\right]
\]
\[\small
m_{22}=\frac{\pi}{2} k_{1} \rho_{0}\left[J_{m}\left(k_{1} \rho_{0}\right) Y_{m}^{\prime}\left(k_{1} \rho_{1}\right)-Y_{m}\left(k_{1} \rho_{0}\right) J_{m}^{\prime}\left(k_{1} \rho_{1}\right)\right]
\]

In the above expressions $p_{1}=\sqrt{\mu_{1} / \varepsilon_{1}}$. It is apparent that the related matrix elements depend on the radii of the two interfaces. For the \textit{i}th layer, the matrix $M_i$ can be acquired by using some replacements including $\rho_{0} \rightarrow \rho_{i-1}$,  $\rho_{1} \rightarrow \rho_{i}$, $k_{1} \rightarrow k_{i}$, and $p_{1} \rightarrow p_{i}$. There are totally $2N$ layers for an $N$-period bilayer periodic photonic crystal. Consequently, there should be $2N$ matrices to establish the total system matrix $M$, which makes a connection between the first and final interfaces as

\begin{equation}\small
\left[ \begin{array}{c}{V\left(\rho_{f}\right)} \\ {U\left(\rho_{f}\right)}\end{array}\right]=M_{2 N} \ldots M_{2} M_{1} \left[ \begin{array}{c}{V\left(\rho_{0}\right)} \\ {U\left(\rho_{0}\right)}\end{array}\right]=M \left[ \begin{array}{c}{V\left(\rho_{0}\right)} \\ {U\left(\rho_{0}\right)}\end{array}\right]
\label{equ8}
\end{equation}

The reflection coefficient of the supposed APC would be obtained by the following equations:

\begin{equation}\small
r_{d}=\frac{\left(M_{21}^{\prime}+j p_{0} C_{m 0}^{(2)} M_{11}^{\prime}\right)-j p_{f} C_{m f}^{(2)}\left(M_{22}^{\prime}+j p_{0} C_{m 0}^{(2)} M_{12}^{\prime}\right)}{\left(-j p_{0} C_{m 0}^{(1)} M_{11}^{\prime}-M_{21}^{\prime}\right)-j p_{f} C_{m f}^{(2)}\left(-j p_{0} C_{m 0}^{(1)} M_{12}^{\prime}-M_{22}^{\prime}\right)}
\end{equation}


In the above expression, $M_{11}^{\prime}$, $M_{12}^{\prime}$, $M_{21}^{\prime}$ and $M_{22}^{\prime}$ are the elements of the inverse matrix of $M$ (see Equation~\ref{equ8}). $p_{0}=\sqrt{\mu_{0} / \varepsilon_{0}}$ and $p_{f}=\sqrt{\mu_{f} / \varepsilon_{f}}$ are the admittances of the initial and last media for the incident wave. $K=\omega \sqrt{\varepsilon_{0} \mu_{0}}$ is the wavenumber in the free space and $C_{m l}^{(1,2)}$ is calculated by the following equation,

\begin{equation}
C_{m l}^{(1,2)}=\frac{H_{m}^{(1,2)}\left(k_{l} \rho_{l}\right)}{H_{m}^{(1,2)}\left(k_{l} \rho_{l}\right)}, \quad l=0, f
\end{equation}

Here, $H_{m}^{(1)}$ and $H_{m}^{(2)}$ are the Hankel function of the first and second kind. The reflectance ($R$) of the APC would be acquired by applying the expression

\[
R=\left|r_{d}\right|^{2}
\]

The formula of reflectance and transmittance for the TM wave would be acquired by using simple replacements including $\varepsilon \leftrightarrow \mu$ and $j \leftrightarrow -j$ in the equation of the TE wave. In the above formulas, the Gorter-Casimir two fluid model \cite{tinkham2004introduction} in the absence of an external magnetic field has been applied to describe the refractive index of the superconductor. A two-fluid model is employed to demonstrate the electrodynamics of a superconductor at non-zero temperature. Based on the related model, the electrons in the superconductor occupy one of the two states, superconducting and normal states. The conductivity of the superconductor is given by
\[
\sigma=\sigma_{1}-j \sigma_{2}
\]
According to the above equation, $\sigma_{1}$ and $\sigma_{2}$ are real and imaginary part respectively and demonstrates the conductivity of normal and superconductor electrons. The real part of the complex conductivity can be ignored for a loss less superconductor, so the conductivity of the superconductor would be expressed as \cite{wu2005photonic, tinkham2004introduction}

\begin{equation}
\sigma=\frac{-j}{\omega \mu_{0} \lambda_{L}^{2}}
\label{equ11}
\end{equation}

Here, $\mu_{0}$ is the vacuum permeability and $\lambda_{L}$ is the temperature-dependent London penetration depth written as

\begin{equation}
\lambda_{L}=\frac{\lambda_{L}(0)}{\sqrt{1-\left(\frac{T}{T_{c}}\right)^{4}}}
\label{equ12}
\end{equation}
where $\lambda_{L}(0)$ is considered as the penetration depth at $T = 0$~K and $T_c$ is the critical temperature of the superconductor. The refractive index of the superconductor would be calculated according to Equation~\ref{equ11} and written as follows:

\begin{equation}
n_{2}=\left(1-\frac{c^{2}}{\omega^{2} \lambda_{L}^{2}}\right)^{1 / 2}
\label{equ13}
\end{equation}

In what follows, we shall calculate the reflection response of a defective annular photonic crystal and discuss the effect of the thickness of layers, the azimuthal mode number, starting radius ($\rho_{0}$), and temperature on the defect modes by studying the reflection spectra.

\section{Numerical result and discussion}

In this work, to calculate the reflection response of the defective APC structure, the alternate layers of the cylinder are taken as semiconductor (silicon monoxide, SiO) with refractive index $n_{1}=1.95$ and high-temperature superconductor Bi$_{2}$Sr$_{2}$CaCu$_{3}$O$_{8}$ (BSCCO) with $T_{c}=95$~K and London penetration depth at $T=0$~K, $\lambda_{L}(0)=150$~nm. A radial defect layer of (Indium nitride, InN) at the middle of the alternate layers of the cylinder with refractive index $n_{3}=2.3$ is considered. The refractive index and London penetration depth of BSCCO at operating temperature $T=4.2~K$ can be acquired by applying Equations~\ref{equ12} and \ref{equ13}, respectively. The thickness of SiO, BSCCO, and InN are considered as $d_{A}=60$~nm, $d_{B}=120$~nm and $d_{C}=20$~nm, respectively, and the total number of periods is $N = 10$. 

\begin{figure}[!b]
\centering\includegraphics[width=0.9\linewidth]{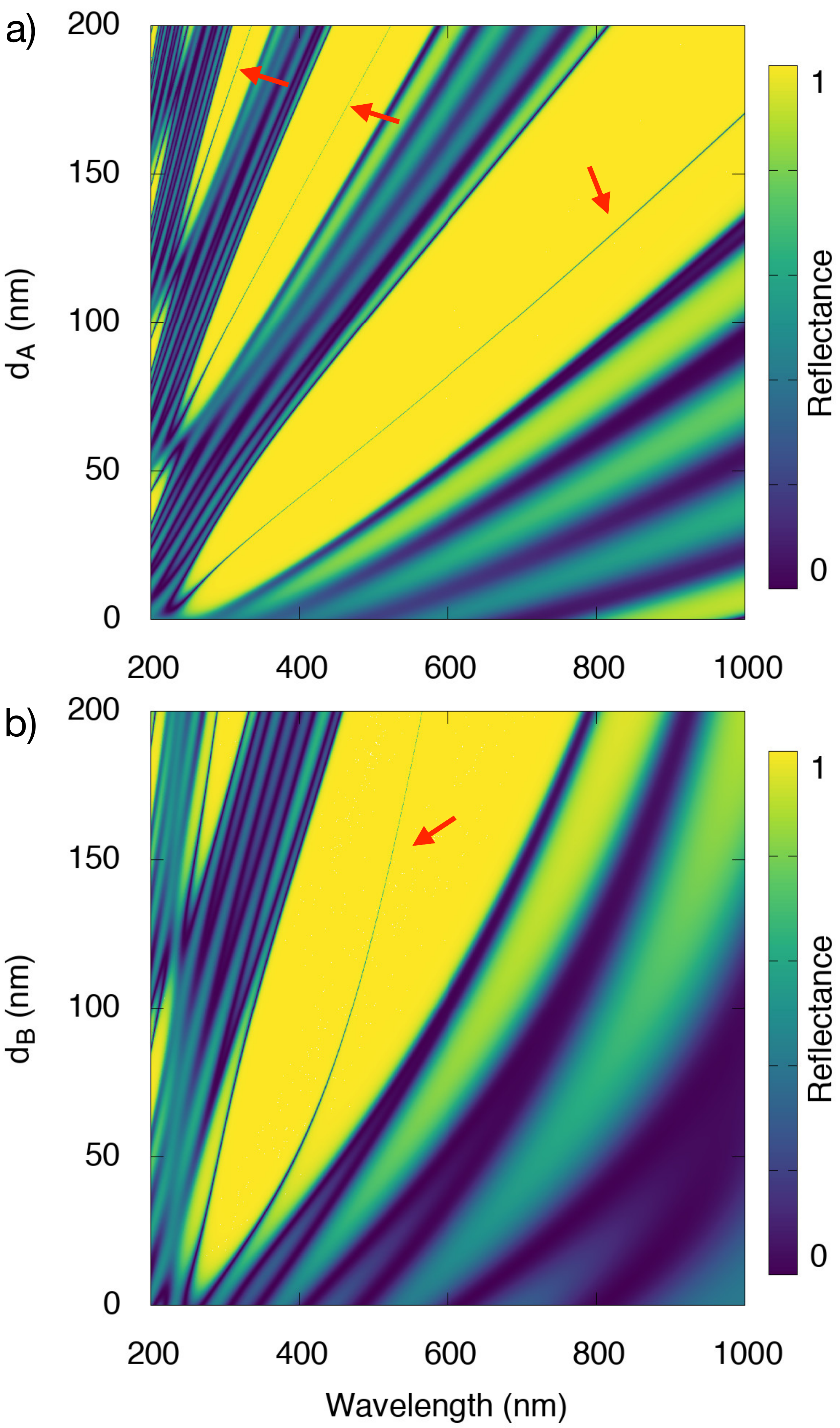}
\caption{ Calculated wavelength-dependent reflectance for a defective APC at $d_{C}=20$~nm, $m=2$ and $\rho_{0}=500$~nm. (a) Depicts the effects of thickness variations of the semiconductor SiO layer. (b) Depicts the effects of thickness variations of the superconductor BSCCO layer. Yellow regions, dark blue regions, and red arrows respectively denote high reflectance rate, zero reflectance (or high transmittance rate), and defect modes. }
\label{fig2}
\end{figure}

First, we investigate the geometric effects of alternate layers on the optical reflectance of electromagnetic waves by modifying the thickness of both SiO and BSCCO layers separately. Hence, the possibility of controlling spectral position of the existed photonic band gaps PBGs and defect modes would be discussed by the following analysis. Figure~\ref{fig2} demonstrates the wavelength-dependent TE-polarized reflectance of the defective APC structure where the starting radius and the azimuthal mode number are respectively considered to be $\rho_{0}=500$~nm and $m=2$, also the thickness of the defect layer and temperature have the constant values of $d_{C}=20$~nm and $T=4.2$~K. There is an indicator for colors at the right hand of all figures which shows the dependency of colors on the reflectance rate. As colors become close to yellow (light) the amount of reflectance increases and as they become close to blue (dark) the amount of transmittance increases

As can be observed in Figure~\ref{fig2}(a), when the thickness of the superconductor layer is kept fixed at 120~nm, by increasing the thickness of the SiO layer, the existed photonic band gaps PBGs (yellow region) and defect modes (denoted with a red arrow) indicate a red-shift trend by moving towards higher wavelengths. Moreover, the increase of $d_A$, leads to an increase in the width and number of the band gaps which appear in the reflectance spectrum. 

Figure~\ref{fig2}(b) depicts the dependence of band gaps and defect mode on the thickness of the superconductor layer when the thickness of the semiconductor layer is set to be constant, $d_{A}=60$~nm. The effect of increasing the thickness of the superconductor layer is similar to that of increasing the thickness of the semiconductor SiO layer as shown in panel (a), which results in increasing the width of the band gap and shifting the defect mode to the higher wavelength values. Based on above results, it is interesting to note that the impact of altering the superconductor layer thickness on the reflectance spectrum is less than that of the semiconductor SiO layer. Another notable result, which may be practical in designing specific optical devices, suggests that by varying the thickness of SiO and BSCCO layers, the band gaps can be adjusted from visible light to near-infrared frequencies.

\begin{figure}[!b]
\centering\includegraphics[width=0.9\linewidth]{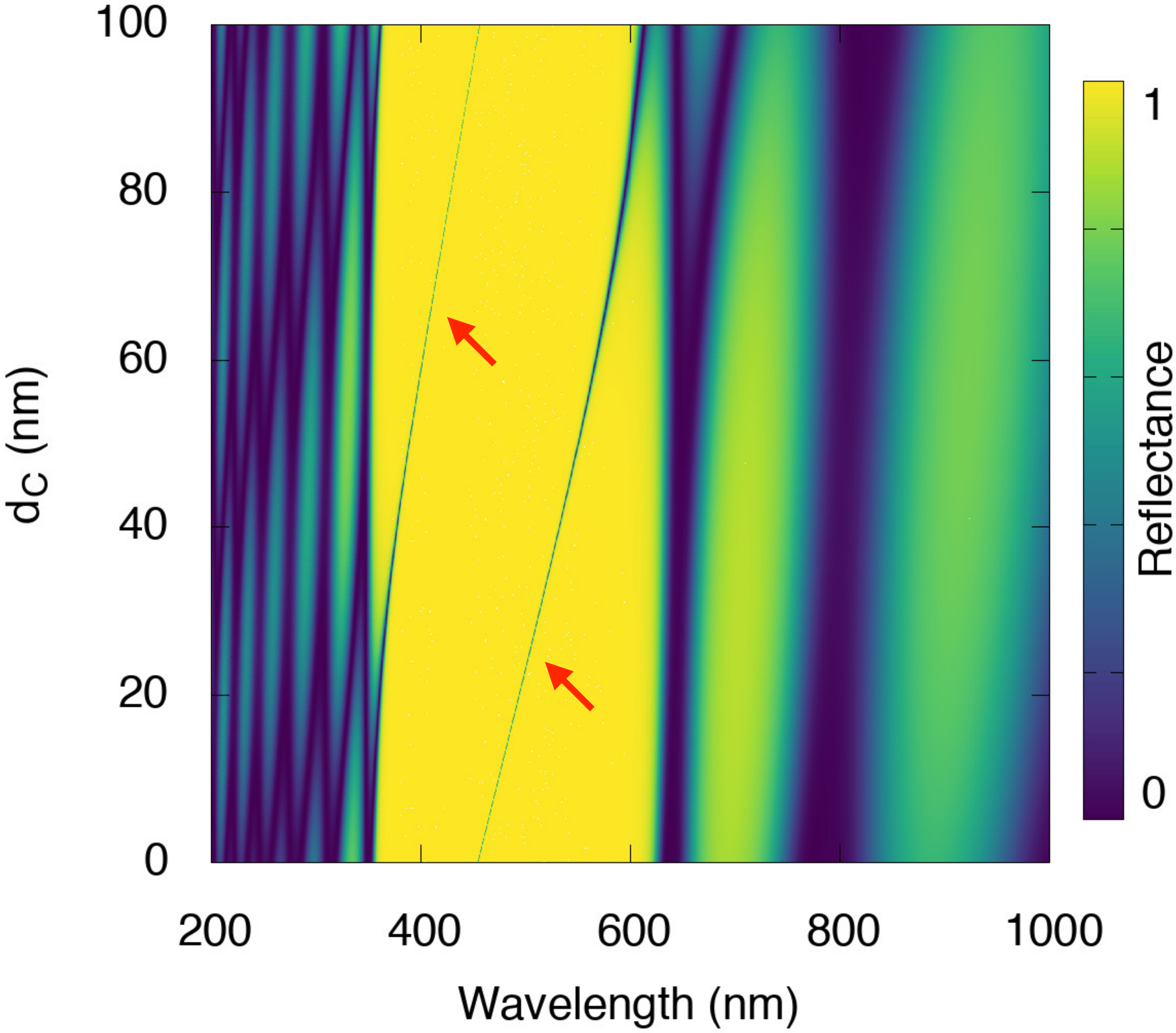}
\caption{ Calculated wavelength-dependent reflectance for an annular PC at $d_{A}=60$~nm, $d_{B}=120$~nm, $m=2$ and $\rho_{0}=500$~nm under the condition of changing the thickness of the defect layer InN. Color coding details are the same as Figure~\ref{fig2}. }
\label{fig3}
\end{figure}

In the following section, we examine the behavior of the band gap and position of the defect modes under the condition of changes in the thickness of the defect layer InN. The reflectance spectrum of the proposed defective APC is plotted in Figure~\ref{fig3}, where $d_{A}=60$~nm, $d_{B}=120$~nm, $m=2$ and $\rho_{0}=500$~nm are considered to be constant. As shown in the figure, increasing the defect layer thickness leads to the movement of the defect mode to higher wavelengths and also another defect mode gradually appears in the band gap. As can be observed from the results, note that a new defect mode with the same pattern as the previous defect mode gently emerges at around $d_{C}=30$~nm in the band gap. We note that there just exists one specified wide band gap. A key point which should be taken into consideration is the position and width of the band gap, which is entirely independent on the thickness of the defect layer. In addition, as shown in Figure~\ref{fig3}, the wavelength of the existed band gap extends from around 350~nm to 620~nm which almost covers the visible spectrum. 

\begin{figure}[!b]
\centering
\includegraphics[width=0.9\linewidth]{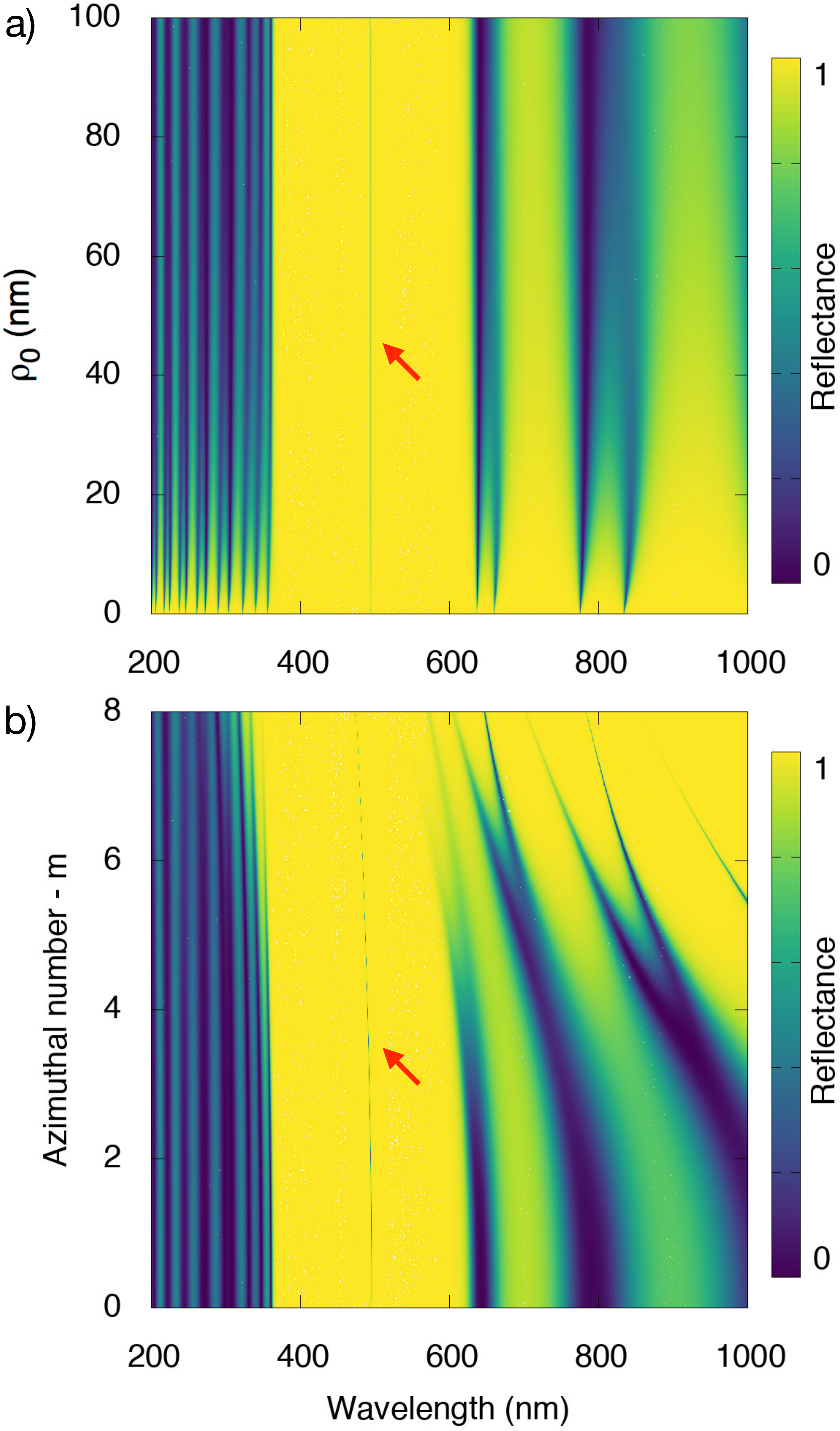}
\caption{ Calculated wavelength-dependent reflectance for a defective APC at $d_{A}=60$~nm, $d_{B}=120$~nm and $d_{C}=20$~nm, and the temperature of the superconductor layer at $T=4.2$~K. (a) Shows the reflectance for various $\rho_{0}$ versus wavelength where $m=2$. (b) Shows the reflectance for various azimuthal mode number $m$ versus wavelength, where $\rho_{0}=500$~nm. Color coding details are the same as Figure~\ref{fig2}.}
\label{fig4}
\end{figure}

Next, we examine separately the effects of both starting radius (the hollow core) of the curved interface $\rho_{0}$ and the azimuthal mode number $m$ on the reflectance characteristics of our design including defect mode and PBGs. Figure~\ref{fig4} depicts the reflectance variations versus wavelength where the thickness of layers is respectively considered to be $d_{A}=60$~nm, $d_{B}=120$~nm and $d_{C}=20$~nm, also the temperature of the superconductor layer is taken to be $T=4.2$~K.

The reflectance for various $\rho_{0}$ as a function of wavelength is examined in Figure~\ref{fig4}(a), where the azimuthal mode number has the value of $m=2$. The figure clarifies that the characteristics of the PBG such as the width of the photonic gap and its spectral position, as well as the position of the defect mode are entirely independent on increasing the starting radius of the curved interface, $\rho_{0}$. In regard to spectral regions, the wavelength of the existed band gap extends from around 360~nm to 620~nm which approximately covers the visible spectrum as shown in panel~(a). On the other side, increasing the size of the core radius (compared to the thickness of alternate layers) can offer flexibility and simplicity of fabrication in producing optical devices. Consequently, starting radius is a parameter of interest from the manufacturers viewpoint.

Figure~\ref{fig4}(b) illustrates the effect of the azimuthal mode number $m$ as a function of wavelength on the optical reflectance, where starting radius of the curved interface is fixed at $\rho_{0}=500$~nm. In this case, it is very interesting to note that as the azimuthal mode number increases, the existed band gaps and defect mode are slightly shifted to lower wavelengths and represent a blue-shift trend by moving toward the blue light wavelength. For instance, take the defect mode at $m=0$ and $m=8$, the defect mode wavelength has the value of around 493~nm at $m=0$ and its value decreases to approximately 471~nm at $m=8$. In addition, it is noteworthy to consider the strong dependence of the band gaps on azimuthal mode number variations as shown in Figure~\ref{fig4}(b). The results show that further band gaps appear at higher wavelengths in $m>4$. The results predict that for higher azimuthal mode numbers, i.e. $m>8$, and except the defect mode, our designed defective APC structure works as a perfect reflector for the considerable range of wavelength.

\begin{figure}[!t]
\centering\includegraphics[width=0.9\linewidth]{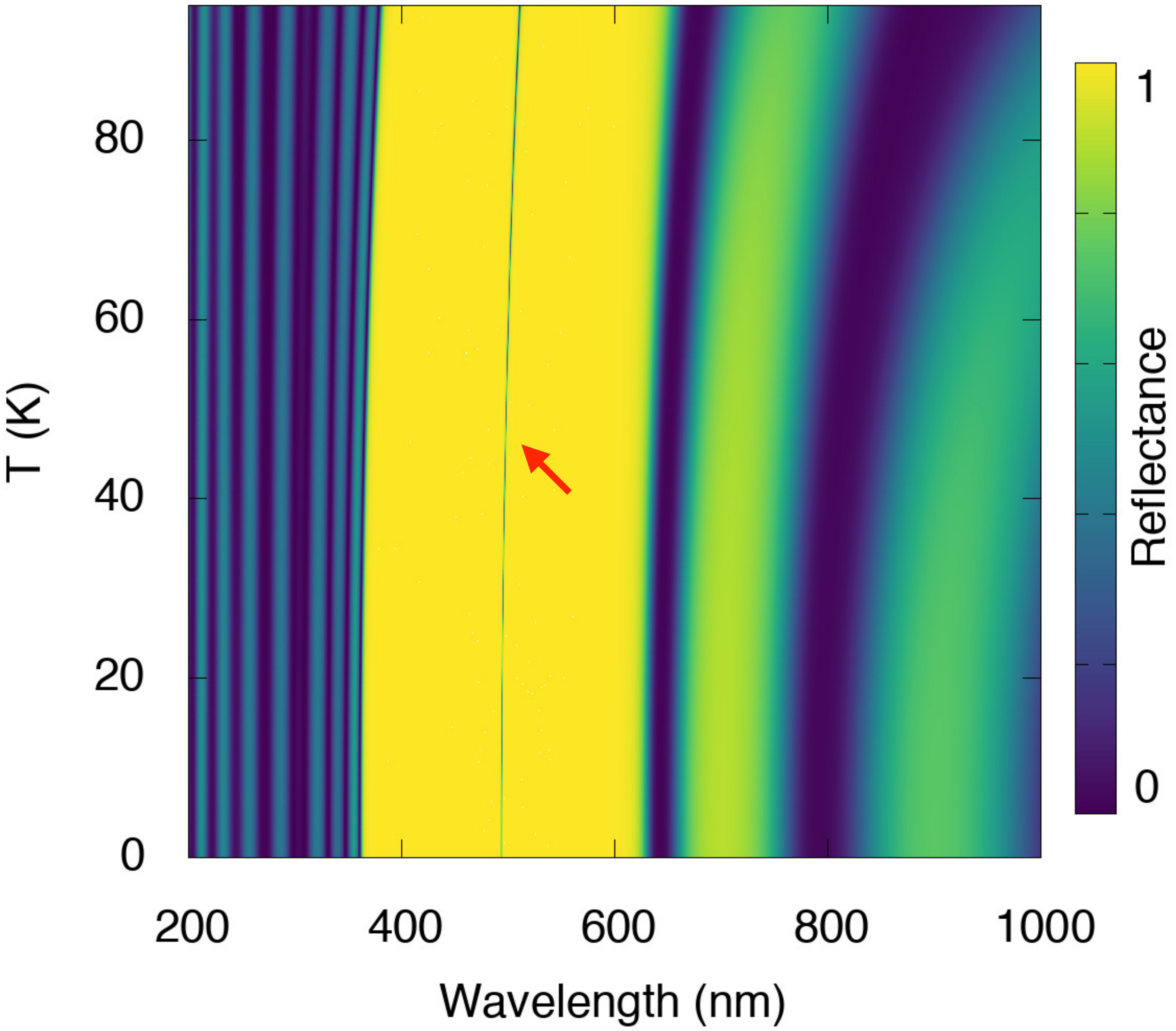}
\caption{ Calculated wavelength-dependent superconductor temperature for a defective APC at $d_{A}=60$~nm, $d_{B}=120$~nm and $d_{C}=20$~nm, where the azimuthal mode number and the starting radius are respectively set to be $m=2$ and $\rho_{0}=500$~nm. Color coding details are the same as Figure~\ref{fig2}. }
\label{fig5}
\end{figure}

In what follows, we investigate the effect of the operating temperature of the superconductor layer on the reflectance spectrum of the proposed defective APC structure. Besides, the dependency of the spectral position of PBGs and defect mode on the superconductor layer temperature is taken into consideration as shown in Figure~\ref{fig5}, where similar to Figure~\ref{fig4}, the thicknesses of layers are considered to be $d_{A}=60$~nm, $d_{B}=120$~nm and $d_{C}=20$~nm. The azimuthal mode number and the starting radius of the curved interface are taken to be $m=2$ and $\rho_{0}=500$~nm, respectively. 

The results indicate that by increasing the superconductor layer temperature, both of the existed band gaps and defect mode show a red-shift trend by moving towards higher wavelengths. On the contrary, the width of the band gap has nearly remained fixed by increasing the temperature. For instance, the wavelength of the defect mode has the value of around 490~nm at $T=10$~K and its value is approximately about 505~nm at $T=90$~K. Before going into the conclusion, it is worth mentioning that the related behavior may be considered attractive for designing temperature sensitive applications which are dependent on the effect of temperature on a monochromatic light. There, such an idea can be put into practice by calibrating changes in the reflectance values against temperature for a specific wavelength of the incident electromagnetic wave. 

\section{Conclusion}
To sum up, we examined theoretically the effects of the thickness of different layers, azimuthal mode number, starting radius, and temperature on the reflectance spectra of defective APCs. The cylindrical designed structure is composed of alternate layers which is interrupted with a radial defect layer. The numerical investigations are demonstrated using the transfer matrix method. According to numerical results, by increasing the thickness of alternate layers, existed PBGs and defect modes indicate a red-shift trend. This also leads to increase in the width of the band gaps which appear in the reflectance spectrum. The spectral position and width of the band gap is entirely independent on the thickness of the defect layer by increasing the thickness of the defect layer. When the starting radius of the curved interface changes, the existed band gap and defect mode are entirely unaffected. It is worthwhile to note that since larger core radius offers simplicity of fabrication in producing optical devices, so this feature is extremely attractive to manufacturers. In addition, there is a strong dependence of the band gaps on azimuthal mode number variations. The results predict that for higher azimuthal mode numbers, our designed defective APC works as a perfect reflector for the considerable range of wavelengths. The results also reveal that by increasing the superconductor layer temperature, both of the existed band gaps and defect mode show a red-shift trend, while the width of the band gap almost remain unaffected by increasing the temperature. The related behavior may be considered attractive for designing temperature sensitive applications which are dependent on the effect of temperature on a monochromatic light. 

\bibliography{References.bib}

\end{document}